# Shape Transition to a Rare Shape Phase of Prolate Non-collective in A=100 Isobars

MAMTA AGGARWAL

Department of Physics, University of Mumbai, kalina Campus, Mumbai 400 098.

*Email: mamta.a4@gmail.com

**Abstract** A theoretical investigation on the shape transitions with neutron number, temperature and spin for A = 100 isobars of Z = 42 to 50 is presented. A variety of shape transitions are observed while moving from neutron rich $^{100}$Mo to proton rich 100Sn with predominant triaxiality. Temperature and spin induced shape transitions are explored within the microscopic theoretical framework of and statistical theory of hot rotating nuclei. Prolate non-collective – a rare shape phase is reported in this mass region on the proton rich side of the nuclear chart.

## 1. INTRODUCTION

Most of the atomic nuclei are known to have spherical or prolate ground state shapes except in few regions on the nuclear chart where oblate and triaxial shapes [23,24,33] are predominant. The nuclear masses around A$\sim$ 100 [7,13-15,17,28,29,32] offer an interesting region to explore shape changes and shape coexistence in the vicinity of Z > 40 and then transition towards less deformed to spherical while approaching shell closure Z = 50. Evidences of triaxial deformation [10,27,30,35] with expected $\gamma$ softness at low spins, speculation of rare shape phase of prolate non-collective for proton rich $^{94}$Ag (Z = 47) [18], shape coexistence [11,16] and shape phase transitions [8,16,25] with changing neutron number, excitation energy and spin provide stringent tests to microscopic models to study shapes and structure of nuclei in this region.

So far it has been well recognized [2,22] that the spin, isospin and temperature have a profound effect on the intrinsic shape of a nucleus and it undergoes a variety of shape transitions with changing neutron number, temperature and angular momentum. The study of such shape-phase transition like phenomena in ground state and excited state nuclei has emerged as a very active field of research in contemporary nuclear physics. Prediction of a rare shape phase of prolate non-collective in one of our recent work [18] in agreement with the speculation of prolate shape by Mukha et al. [12]
opened up new avenues to explore domains of periodic table with states of such rare shape phases. This unexpected prolate non-collective equilibrium phase first anticipated by Goodman [4] and then predicted by us for proton rich $^{94}$Ag (Z = 47) generated directly by rotation at certain angular momentum values undergoes the expected transition to the oblate non-collective phase at higher angular momentum values [18]. To explore such rare shape phases in nuclei around Z = 47 and shape changes induced by the neutron number, temperature and rotation in A$\sim$ 100 [[17] region is the objective of present work. We focus, in particular, on nuclei ranging from neutron rich to neutron deficient nuclei and investigate in a theoretical framework the nature and the consequences of structural transitions in A=100 (Z = 42 to 50) isobars in ground state as well as in low excitation state at various spins.

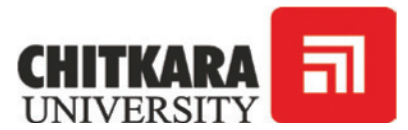

In nuclei (Z, N < 40), ground state prolate and oblate shapes are found to occur more or less equally but for N, Z > 50 the prolate shapes are more probable [9] although there have been evidences of triaxial deformation in Z > 40 [10,27,30,35] with expected $\gamma$ softness at low spins. Oblate shapes are expected to occur just below the N = 82, 126 and Z = 82 shell closures due to the strong shape-driving effect of holes in the $\Omega$ = 1/2 orbitals [1]. One may also find configurations corresponding to different shapes coexisting [16] at similar energies as shown in our recent works [19-20] for odd Z rare earth proton rich nuclei (Z = 51 to 75) along with the shape transition from the unusual prolate non-collective shape phase to usual oblate non collective shape phase which was found absent in even Z Te isotopes [21] although they all show rapid shape transitions from usual shape phases of prolate to triaxial and oblate.

The theoretical tools for the analysis of nuclear shape transitions are usually based on some mean field approximations [9] for the ground state nuclei and finite temperature Nilsson-Strutinsky Cranking method used by the Dubna group [3], FTHFB used by Goodman [5], Landau theory by Alhassid et al [36] and statistical theory of hot rotating nuclei by our group initiated by Rajasekaran et al. [22] for excited nuclei. Although mean field theories are being used to understand structural transitions in nuclei, our formalism using the triaxially deformed Nilsson potential and Strutinsky's prescription for ground states combined with statistical theory [18,20] of hot rotating nuclei to investigate excited states provides in its own way a simplistic and effective approach to trace structural aspects of ground [19] as well as excited nuclei [21].

## 2. THEORETICAL FORMALISM (SECTION 2)

Nuclear shapes are governed by delicate interplay of the macroscopic bulk properties of nuclear matter and the microscopic shell effects which are treated with Nilsson-Strutinsky prescription which starts with the well known Strutinsky density distribution function [34] for single particle states (See Ref. [19-20] for detailed formalism). Strutinsky's shell correction $\delta E_{Shell}$ added to macroscopic energy of the spherical drop $BE_{LDM}\{PM\}$ along with the deformation energy $E_{def}$ obtained from surface and coulomb effects gives the total energy $BE_{gs}$ as in our earlier works [18,21] corrected for microscopic effects of the nuclear system.

$$BE_{gs}(Z,N,\beta,\gamma) = BE_{LDM}(Z,N) - E_{def}(Z,N,\beta,\gamma) - \delta E_{Shell}(Z,N,\beta,\gamma) \quad (1)$$

Energy E (= -BE) minima are searched for various $\beta$ (0 to 0.4 in steps of 0.01) and $\gamma$ (from -180$^o$ (oblate) to -120$^o$ (prolate) and -180$^o$ < $\gamma$ < -120$^o$ (triaxial)) to trace the nuclear shapes and equillibrium deformations.

To evaluate deformation and shape of the excited nucleus we calculate excitation energy E* and entropy S of the hot rotating nuclear system for fixed T and M given as an input as a function of $\beta$ and $\gamma$ and then incorporate them to the ground state energy calculated using macroscopic-microscopic approach [19] and then minimize free Energy (F) with respect to deformation parameters ($\beta$, $\gamma$) at temperature T and angular momentum M.

$$\begin{aligned} F(Z,N,T,M,\beta,\gamma) = {} & E_{LDM}(Z,N) + \delta E_{Shell}(\beta,\gamma) + E_{def}(Z,N,\beta,\gamma) \\ & + E^{*}(T,M,\beta,\gamma) - TS(T,M,\beta,\gamma) \end{aligned} \quad (2)$$

The Free energy minima gives the deformation and shape of the hot rotating nucleus. We use the convention for the axial deformation parameter $\beta$ from 0 to 0.4 in steps of 0.01 and the angular deformation parameter $\gamma$ ranges from -180$^o$ (oblate with symmetry axis parallel to the rotation axis) to -120$^o$ (prolate with symmetry axis perpendicular to rotation axis) and then to -60$^o$ (oblate collective) to 0$^o$ (prolate non-collective). The calculations are performed for A=100 isobars for Z = 42 to 50 nuclei. The value of temperatures used is very low so as to understand the effect of rotation on nearly yrast states which may not be seen at high temperatures because the nuclear deformation is caused by quantum shell effects and increasing the temperature creates thermal excitations which eventually wash out these shell effects driving the equilibrium shape to spherical [6] and the nucleus resembles a classical liquid drop.

## 3. RESULTS AND DISCUSSION (SECTION 3)

Here we present the results of our theoretical calculation in the microscopic framework for A=100 isobars ranging from neutron rich $^{100}Mo_{58}$ to neutron deficient $^{100}Sn_{50}$ nuclei that provide an ideal testing ground for the observation of various nuclear structural phenomena with changing neutron number, spin and temperature. Figure 1(a). shows the variation of equilibrium deformation and shape parameters ($\beta$, $\gamma$) of A=100 isobars of Z = 42 to 50. Our calculated values of $\beta$ agree very well with the available experimental data [31] and the other theoretical data [26]. Triaxial shape with large deformation ($\beta$=0.24 to 0.17) is found to be predominant in A=100 region as also identified by Moller [26] for ground state nuclei. As N decreases, we find ground state energy minima shifts towards prolate shape with decreasing $\beta$ while moving towards shell closure at neutron deficient $^{100}Sn$ (See Fig. 1(b)) where equilibrium deformation becomes zero.

Ground state $\gamma$-soft energy minima depicting shape and deformation is shown in Fig. 2 where E is plotted vs. ($\beta$, $\gamma$). Shape transition from $\gamma$ =

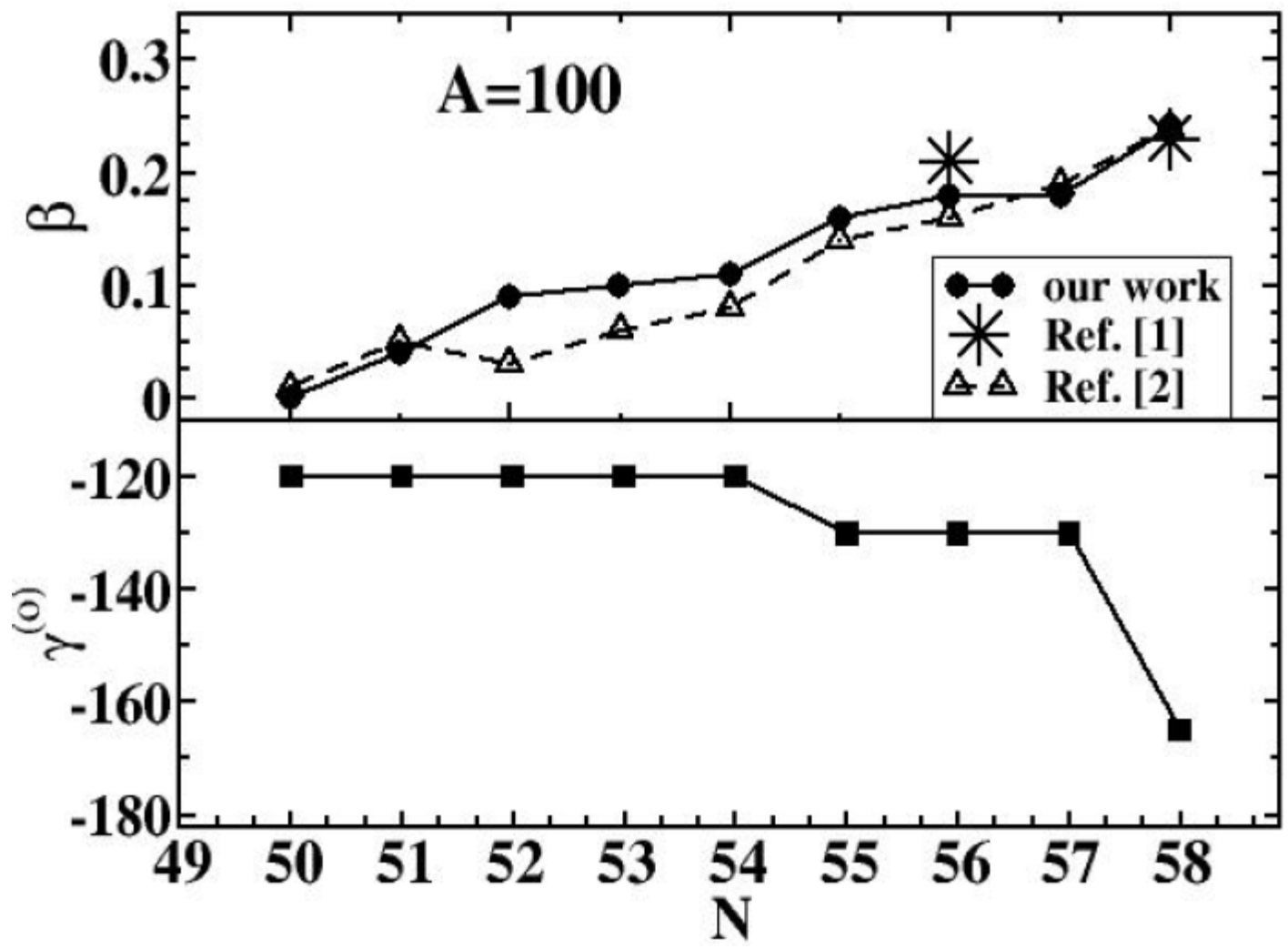

**Figure 1:** (a) Ground state deformation $\beta$ vs. neutron number N for A=100 isobars (b) Shape parameter $\gamma$ vs. N for A=100 isobars.

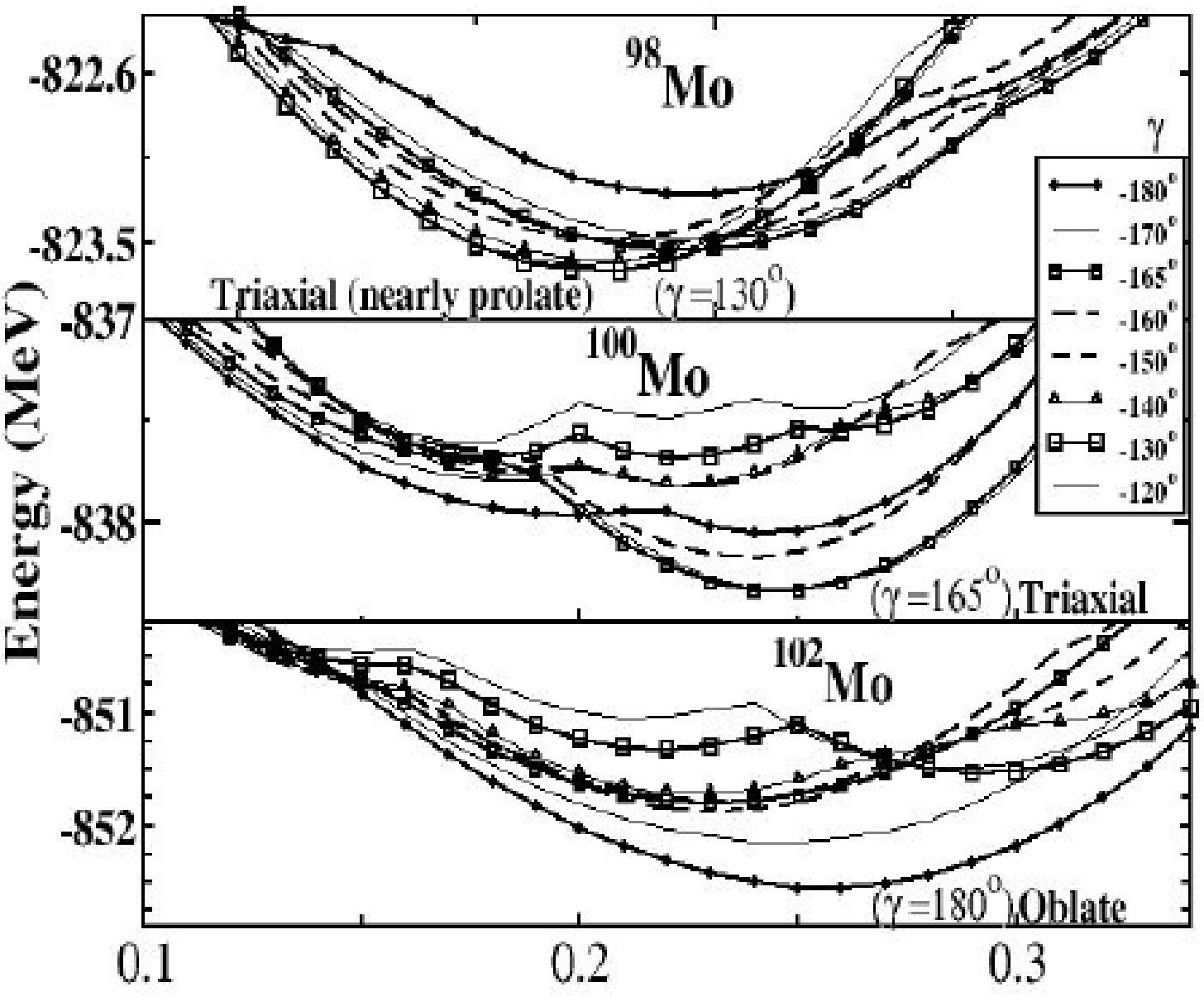

**Figure 2:** Energy minima curves as a function of ($\beta$ $\gamma$) for (a) $^{98}$Mo, (b) $^{100}$Mo and (c) $^{102}$Mo evaluating deformation and shape. Nuclei undergoing shape transition with increasing N is seen.

130$^{o}$(triaxial (nearly prolate), $\beta$=0.17)) at $^{98}$Mo to $\gamma$ = 165$^{o}$ (Triaxial, $\beta$ = 0.24) for $^{100}$Mo to $\gamma$ = 180$^{o}$(Oblate, $\beta$=0.26) at neutron rich $^{102}$Mo. Here large deformation is seen with a well defined single minima indicating absence of shape coexistence in these nuclei unlike the two energy minima we saw in $^{113}$Cs [19] and in excited states of $^{131}$Eu [20] in our recent work [19-20] indicating shape coexistence.

A hot rotating nucleus which is an isolated many-body system with a complex structure and study of its intrinsic structural properties undergoing various transitions with changing temperature and rotation has emerged as an active topic of research. At a given temperature and spin, the equilibrium rotational configuration minimizes the nuclear free energy and one finds the evolution of equilibrium shape as a function of temperature and spin [2,4,21,22,36]. We investigate excited states in the region of A=100 isobars for Z = 42 - 50, in the theoretical framework of statistical theory of hot rotating nuclei. For a given temperature T = 0.7 and 1 MeV and spin 0 – 60$\hbar$, we evaluate shape and deformation of $^{100}$Mo-Sn that range from neutron rich to neutron deficient nuclei. Figure 3 and 4 plots $\beta$ and $\gamma$ vs. angular momentum M($\hbar$) at T = 0.7 MeV respectively. Neutron rich nuclei Mo, Tc and Ru have triaxial deformation with $\beta \geq 0.2$ at very low spins with shape transition to oblate non-collective with increasing spin at $M \geq 10\hbar$ whereas the proton rich nuclei Pd, Ag, Cd, In and Sn show the rare shape phase of prolate non-collective (n-c) with $\gamma=0^{o}$ at low and medium spin with a shape transition to oblate shape with increasing spin. It is evident from the figure that prolate non-collective shape phase is the predominant phase on the proton rich side that is Ag, Cd, In and Sn nuclei with transition to usual shape phase of Oblate n-c at M $\geq$ 30$\hbar$ whereas Rh and Pd have prolate n-c shape phase only for a very narrow spin window of around 0-10$\hbar$ for Rh and 0-20$\hbar$ for Pd. With increasing temperature T = 1.0 MeV, equilibrium deformation decreases for all the nuclei studied here. For higher temperatures T≈1.5-2 MeV, equilibrium deformation reduces to zero with nuclear shape moving to sphericity due to melting away of shell effects as shown in our earlier works [20]. We also note in Figures (3) and (5) that the equillibrium deformation shows a dip during the transition from one shape phase of prolate n-c to another shape phase of oblate n-c and then increases to much larger values. It is interesting to note that with increasing angular momentum, the doubly closed shell nuclei $^{100}$Sn also undergoes a shape transition from spherical to well deformed oblate states with $\beta \approx 0.11$.

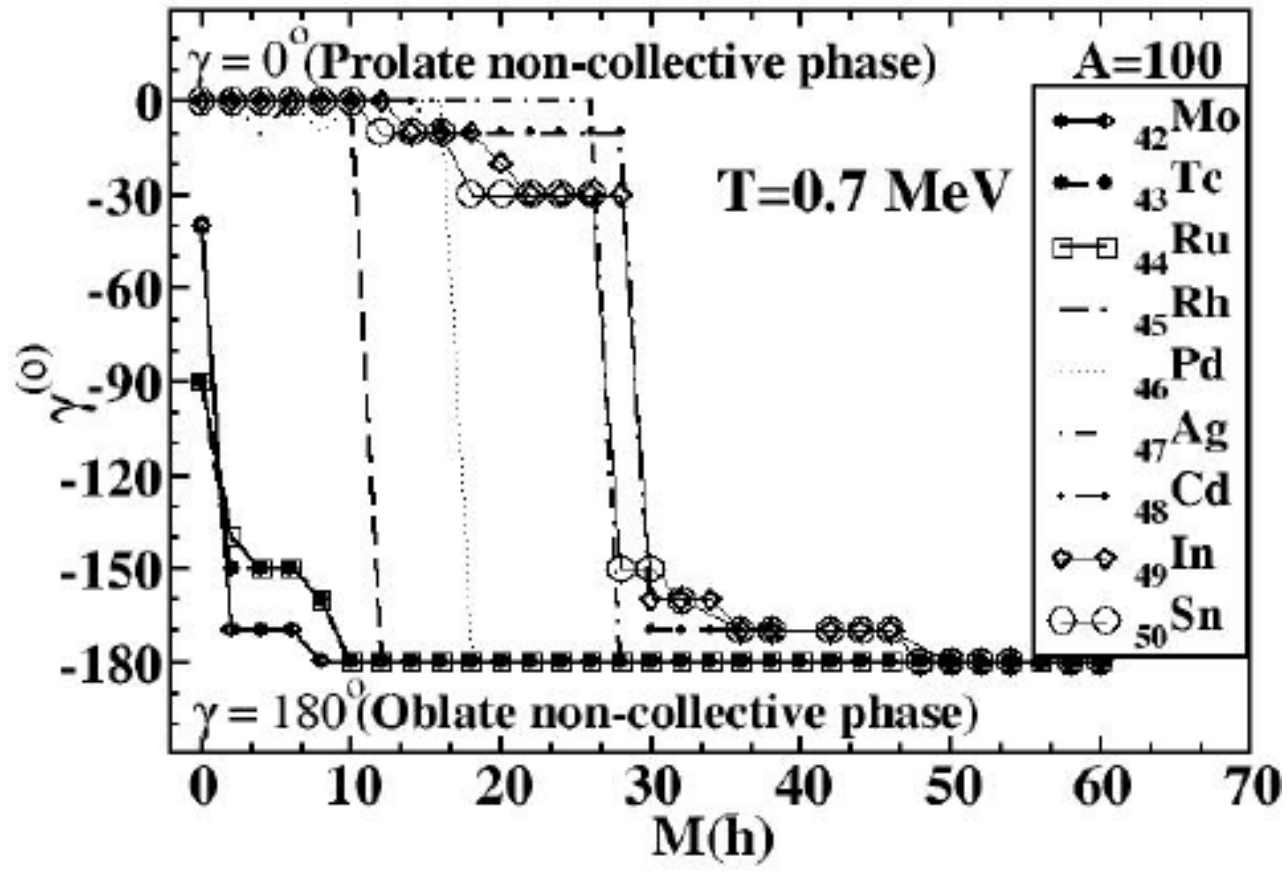

**Figure 3:** Shape parameter $\gamma$ variation with angular momentum M ($\hbar$) showing different shape phases and transition from rare shape phase of prolate n-c to oblate n-c at T= 0.7 MeV for A=100 isobars for Z=42 to 50.

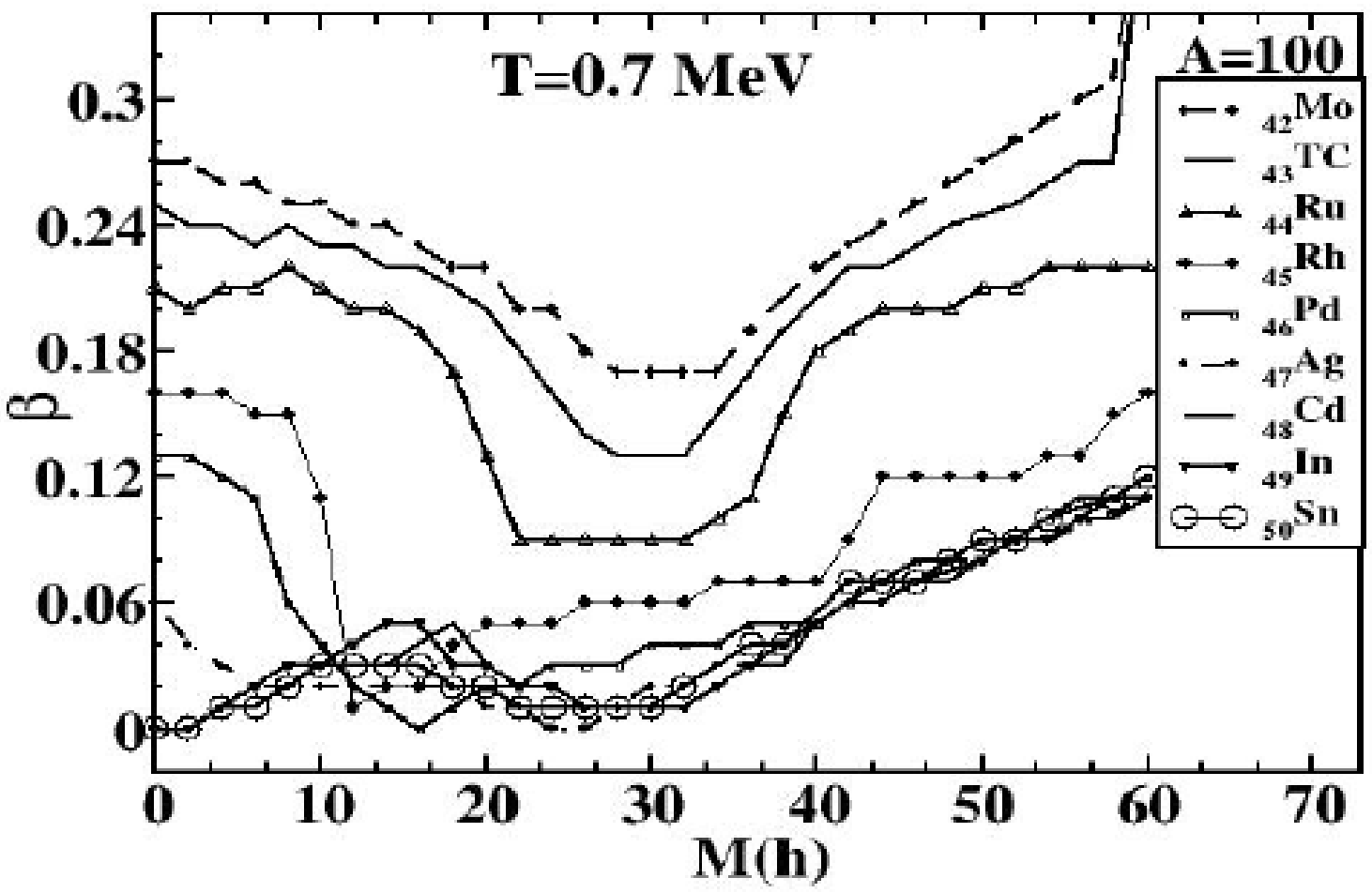

**Figure 4 :** Variation of $\beta$ with angular momentum M($\hbar$) for A=100 isobars for Z = 42 to 50 at T = 0.7 MeV.

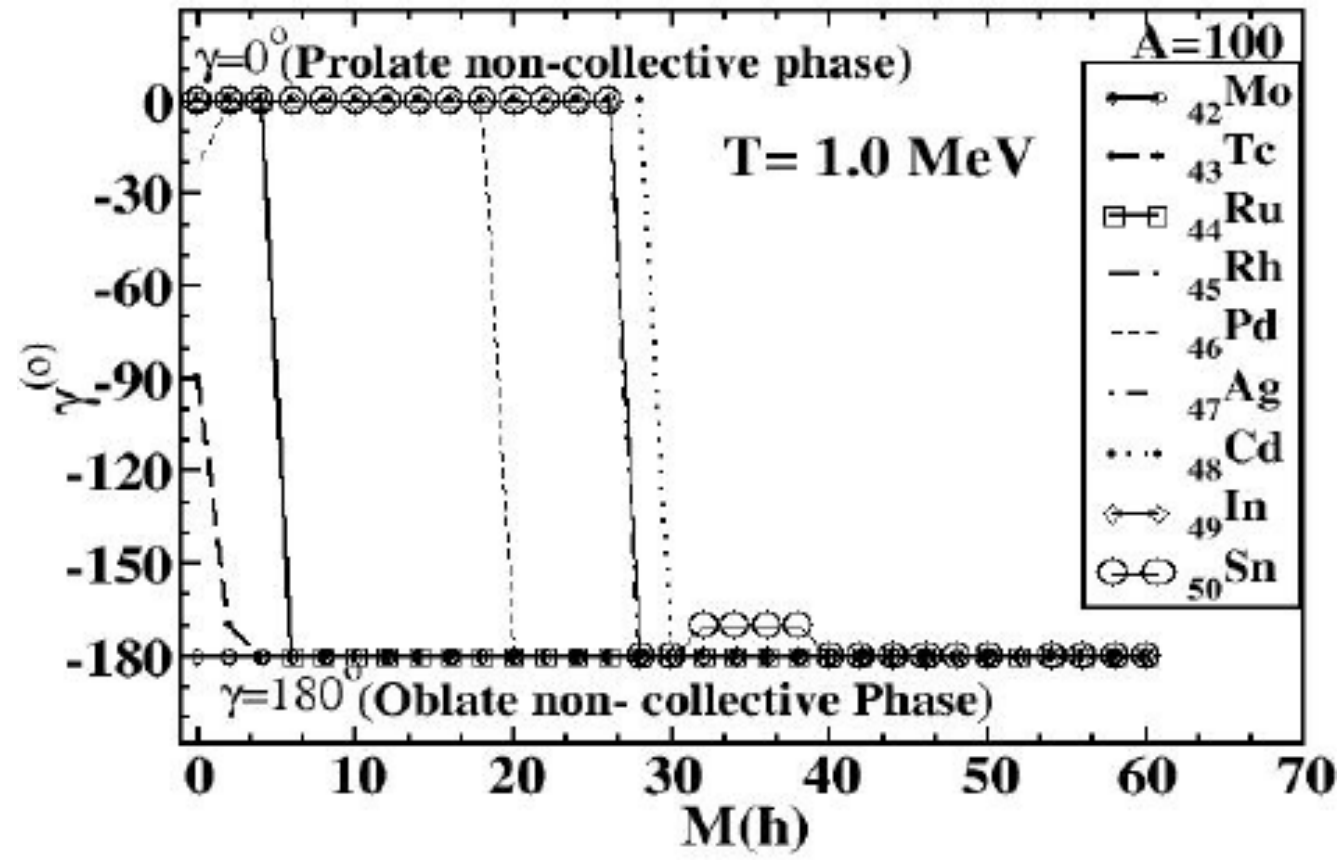

**Figure 5:** Shape parameter $\gamma$ variation with angular momentum M ($\hbar$) showing different shape phases and transition from rare shape phase of prolate n-c to oblate n-c at T= 1 MeV for A=100 isobars for Z=42 to 50.

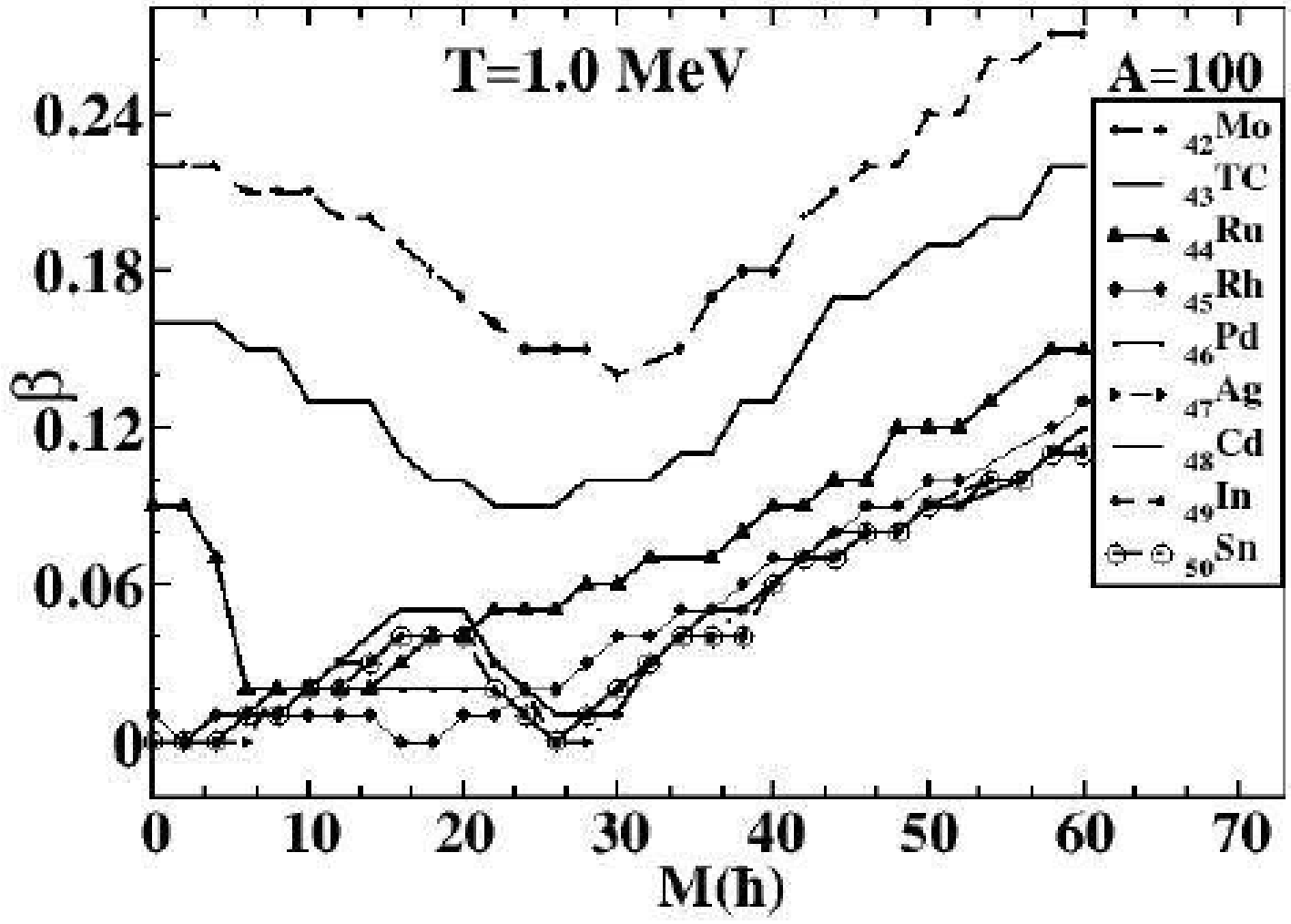

**Figure 6:** Fig. 4 : Variation of $\beta$ with angular momentum M($\hbar$) for A = 100 isobars for Z = 42 to 50 at T= 1 MeV.

It is important to emphasize here that the rare shape phase of prolate n-c is more often seen on the proton rich side of nuclear chart as indicated in our earlier works on rare earth odd Z proton emitters [20] and proton rich nuclei $^{94}$Ag [18] where we showed that this shape phase favours proton radioactivity which had agreed with the experimental work of Mukha [12]. So far this shape phase has been seen in proton emitters in our works although there is no taboo on finding this shape phase on neutron rich side but so far our investigation indicates that proton rich nuclei or rather neutron deficient nuclei show this rare shape phase. Moreover, no other works have reported this shape phase in A=100 region other than our work. We also find that A=100 region exhibits triaxiality as also predicted by Moller [26] and other works [10,27,30,35] in ground state, but we find it only on the neutron rich side of the nuclei whereas proton rich nuclei show prolate shape in the ground state with a shape transition to prolate non-collective state at low excitation and low spin and eventually move to commonly seen shape phase of oblate n-c state at higher spin. This work has lot more potential as it needs much more insights into our predictions with lot more support from experimental data on low spin low excitation states.

## CONCLUSION

Structural aspects of A = 100 isobars ranging from neutron rich side (Z = 42 to 45) to neutron deficient side (Z = 46-50) are studied. In ground state, triaxiality is predominant shape phase with high deformation on the neutron rich side with shape transition to prolate deformations while moving towards proton rich side while approaching shell closure Z = 50 where equilibrium deformation reduces to zero. Coexisting states with similar energies are not seen in the present study. Excited states of rotating nuclei are studied within theoretical framework of statistical theory. We predict for the first time the rarely seen shape phase of prolate non-collective in A=100 isobars on the neutron deficient side of the nuclei at low spin with a shape transition to usually seen shape phase of oblate non-collective. A = 100 isobars of Mo, Tc, Ru do not exhibit this rare shape phase whereas Rh, Pd, Ag, Cd, In and Sn nuclei are predicted to have prolate n-c shape as the predominant shape phase over a large angular momentum range.

## ACKNOWLEDGEMENTS

The author acknowledges financial support from DST under WOS-A scheme. I thanks Dr. S. Kailas for his support and useful discussions for this work.